\documentclass[a4paper,11pt]{article}
\usepackage{jinstpub} 

\usepackage{amssymb}
\usepackage{graphicx}
\usepackage{caption}
\usepackage{subcaption}
\usepackage{siunitx}
\usepackage{lineno}
\usepackage{tabularx}
\usepackage{wasysym} 
\usepackage{hyperref}
\usepackage{url}
\newcommand{\added}[1]{#1}                    

\title{\boldmath Design and Performance of a 96-channel Resistive PICOSEC Micromegas Detector for ENUBET}

\author[a,1]{A. Kallitsopoulou,\note{Corresponding author}}
\author[a]{S.Aune}
\author[c,d]{Y.Angelis}
\author[a]{R. Aleksan}
\author[a]{A. Bonenfant}
\author[d]{J. Bortfeldt}
\author[e]{F. Brunbauer}
\author[f,g]{M. Brunoldi}
\author[h]{J. Datta}
\author[a]{D. Desforge}
\author[i]{G. Fanourakis}
\author[f,g,2]{D. Fiorina\note{Now at Gran Sasso Science Institute, Viale F. Crispi, 7 67100 L'Aquila, Italy}}
\author[e,j]{K. J. Floethner}
\author[k]{M. Gallinaro}
\author[l]{F. Garcia}
\author[a]{I. Giomataris}
\author[m]{K. Gnanvo}
\author[a,3]{F.J. Iguaz\note{Now at SOLEIL Synchrotron, L'Orme des Merisiers, 91190 Saint Aubin, France}}
\author[e]{D. Janssens}
\author[a]{F. Jeanneau}
\author[a]{M.Kebbiri}
\author[n]{M. Kovacic}
\author[h]{B. Kross}
\author[a]{P. Legou}
\author[e]{M. Lisowska}
\author[p]{J. Liu}
\author[a]{C.Loiseau}
\author[k,q]{M. Lupberger}
\author[c,d,4]{I. Maniatis\note{Now at Department of Particle Physics and Astronomy, Weizmann Institute of Science, Rehovot, Israel.}}
\author[n]{J. McKisson}
\author[a]{B.Moreno}
\author[p]{Y. Meng}
\author[f,q]{H. Muller}
\author[f]{E. Oliveri}
\author[f,r]{G. Orlandini}
\author[n]{A. Pandey}
\author[a]{T. Papaevangelou}
\author[s]{M. Pomorski}
\author[a]{E.F.Ribas}
\author[f]{L. Ropelewski}
\author[c,d]{D. Sampsonidis}
\author[f]{L. Scharenberg}
\author[f]{T. Schneider}
\author[s]{E. Scorsone}
\author[b,5]{L. Sohl\note{Now at TUV NORD EnSys GmbH \& Co. KG.}}
\author[f]{M. van Stenis}
\author[t]{Y. Tsipolitis}
\author[c,d]{S.E. Tzamarias}
\author[u]{A. Utrobicic}
\author[f,g]{I. Vai}
\author[f]{R. Veenhof}
\author[g,h]{P. Vitulo}
\author[p]{X. Wang}
\author[v]{S. White}
\author[n]{W. Xi}
\author[p]{Z. Zhang}
\author[p]{Y. Zhou}

\affiliation[a]{IRFU, CEA-Université Paris-Saclay, F-91191 Gif-sur-Yvette, France}
\affiliation[b]{Department of Physics, Aristotle University of Thessaloniki, GR-54124 Thessaloniki, Greece}
\affiliation[c]{CIRI-AUTH, GR-57001 Thessaloniki, Greece}
\affiliation[d]{Department for Medical Physics, Ludwig Maximilian University of Munich, 85748 Garching, Germany}
\affiliation[e]{CERN, 1211 Geneva 23, Switzerland}
\affiliation[f]{Dipartimento di Fisica, Università di Pavia, 27100 Pavia, Italy}
\affiliation[g]{INFN Sezione di Pavia, 27100 Pavia, Italy}
\affiliation[h]{Department of Physics and Astronomy, Stony Brook University, NY 11794-3800, USA}
\affiliation[i]{Institute of Nuclear and Particle Physics, NCSR Demokritos, GR-15341 Agia Paraskeui, Attiki, Greece}
\affiliation[j]{Helmholtz-Institut für Strahlen- und Kernphysik, University of Bonn, 53115 Bonn, Germany}
\affiliation[k]{Laboratório de Instrumentação e Física Experimental de Partículas (LIP), Lisbon, Portugal}
\affiliation[l]{Helsinki Institute of Physics, University of Helsinki, FI-00014 Helsinki, Finland}
\affiliation[m]{Jefferson Lab, Newport News, VA 23606, USA}
\affiliation[n]{Faculty of Electrical Engineering and Computing, University of Zagreb, 10000 Zagreb, Croatia}
\affiliation[o]{State Key Laboratory of Particle Detection and Electronics, University of Science and Technology of China, 230026 Hefei, China}
\affiliation[p]{Physikalisches Institut, University of Bonn, 53115 Bonn, Germany}
\affiliation[q]{Friedrich-Alexander-Universität Erlangen-Nürnberg, 91054 Erlangen, Germany}
\affiliation[r]{CEA-List, Diamond Sensors Laboratory, CEA-Saclay, F-91191 Gif-sur-Yvette, France}
\affiliation[s]{National Technical University of Athens, Athens, Greece}
\affiliation[t]{Ruđer Bošković Institute, 10000 Zagreb, Croatia}
\affiliation[u]{University of Virginia, Virginia, USA}

\emailAdd{alexandra.kallitsopoulou@cea.fr}

\abstract{

The PICOSEC-Micromegas (PICOSEC-MM) detector is a fast gaseous detector concept that achieves picosecond-level timing. It couples a Cherenkov radiator, typically an $\mathrm{MgF_2}$ crystal, to a Micromegas-based photodetector equipped with a photocathode, allowing the fast photoelectron-induced signal to suppress the intrinsic time jitter characteristic of gaseous detectors. This design enables sub-20\,ps timing precision while preserving the robustness and scalability of Micro-Pattern Gaseous Detector (MPGD) technologies. The 96-pad PICOSEC-MM detector represents the latest advancement in this development, optimized for precision timing in high-energy physics. Building upon the R\&D insights obtained with earlier 7-pad resistive prototypes, this large-area demonstrator was developed to validate the scalability, uniformity, and robustness of the technology for integration into the ENUBET project. The detector employs a 2.5\,nm Diamond-Like Carbon (DLC) photocathode with a Micromegas board equipped with surface resistivity of 10~M$\Omega$/\(\square\), providing an excellent timing performance. The prototype was characterized using 150~GeV/$c$ muons at the CERN SPS beamline, with one-third of the active area instrumented during each run. A dedicated alignment procedure, developed for multi-pad PICOSEC-MM systems, was used to reconstruct the pad centers and combine measurements across different detector regions. The measured timing resolution was 43\,ps across the instrumented pads, while the Signal Arrival Time (SAT) distributions exhibited a good uniformity among the detector area that was tested. Mechanical flatness was identified as a key factor influencing timing precision. Maintaining a planarity tolerance within 10~\(\upmu\)m is therefore critical to preserve a good timing resolution over large active areas. The successful operation of the 96-pad demonstrator confirms the scalability of the PICOSEC-MM concept marking a significant step toward implementing robust, high-granularity, picosecond-level gaseous timing detectors in monitored neutrino beam experiments such as ENUBET. }


\begin{document}
\maketitle
\flushbottom

\section{Introduction}\label{sec:introduction}

The next generation of particle physics experiments demands detectors capable of providing sub-100\,ps timing precision while maintaining high-rate capability and radiation hardness. Managing pile-up density---the overlap of multiple particle interactions in time and space---remains one of the most significant challenges in modern high-luminosity facilities~\cite{Detector:2784893}. Precise timing information not only improves triggering and event reconstruction but also enables accurate particle identification and vertex separation in dense environments. Achieving these goals requires detector systems that combine excellent timing performance, large-area coverage, and robust operation under intense particle flux.

\begin{figure}[hbt!]
  \centering
  \begin{subfigure}[b]{0.5\textwidth}
    \centering
    \includegraphics[width=\textwidth]{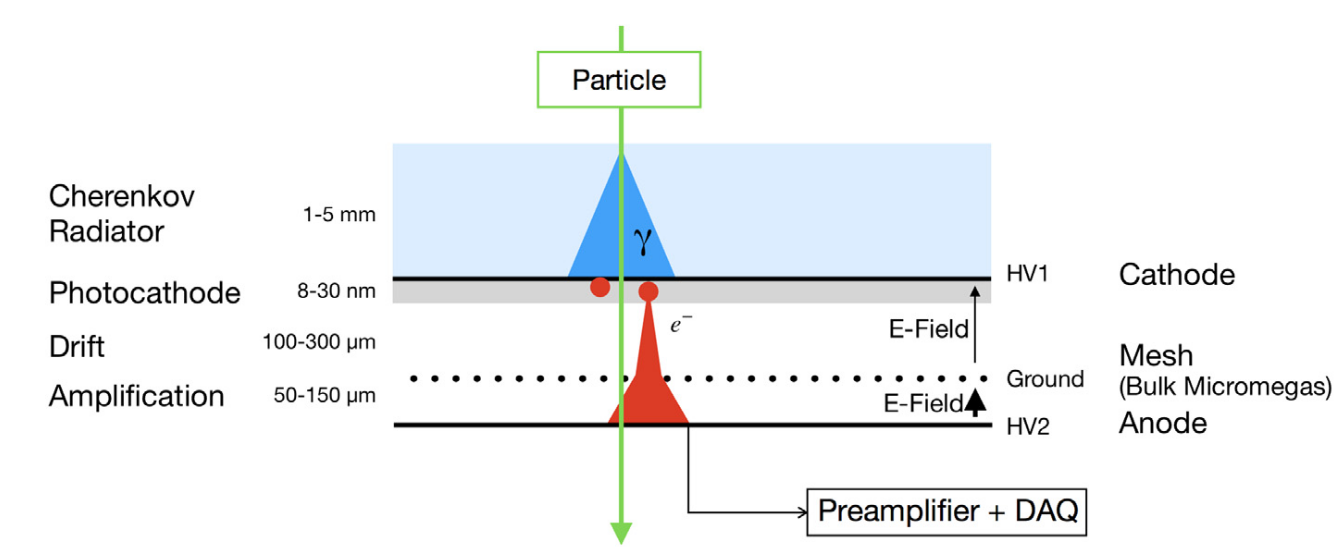}
    \caption{}
    \label{fig:picosec}
  \end{subfigure}
  \hfill
  \begin{subfigure}[b]{0.4\textwidth}
    \centering
    \includegraphics[width=\textwidth]{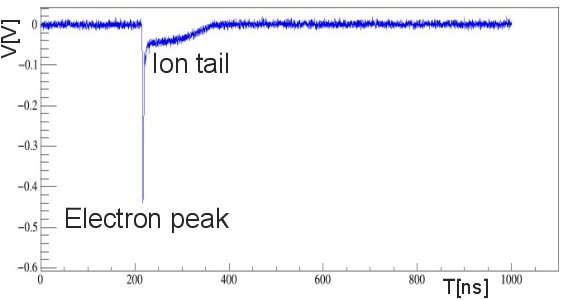}
    \caption{}
    \label{fig:picosec_waveform}
  \end{subfigure}
  \caption{(a) Graphical representation of a PICOSEC Micromegas detector \cite{test}. (b) Typical PICOSEC Micromegas signal after the amplifier.}
\end{figure}

The PICOSEC Micromegas (PICOSEC-MM) technology represents a breakthrough in gaseous detector timing, improving the resolution of standard MicroPattern Gaseous Detectors (MPGDs) by nearly three orders of magnitude \cite{test}. It operates by coupling a Cherenkov radiator and a photocathode to a Micromegas amplification stage. A relativistic charged particle traversing the 3\,mm thick $\mathrm{MgF_2}$ Cherenkov radiator emits photons as its velocity exceeds the speed of light in the medium. These photons strike the photocathode, releasing photoelectrons in a highly synchronized way. Entering the drift region under a strong electric field, these photoelectrons immediately trigger charge multiplication, forming an avalanche. As the avalanche passes through the micromesh into the amplification gap, a higher electric field induces further multiplication, generating a measurable current on the anode that is recorded by the readout electronics (Fig. \ref{fig:picosec}). This mechanism reduces the timing jitter---typically a few nanoseconds in conventional Micromegas---to the tens-of-picoseconds level, enabling unprecedented precision. The characteristic PICOSEC-MM signal consists of a fast ($\sim$700\,ps) electron peak followed by a slower ($\sim$100\,ns) ion tail (see Fig.~\ref{fig:picosec_waveform}).

\begin{figure}[hbt!]
 \centering 
 \includegraphics[width=0.5\textwidth]{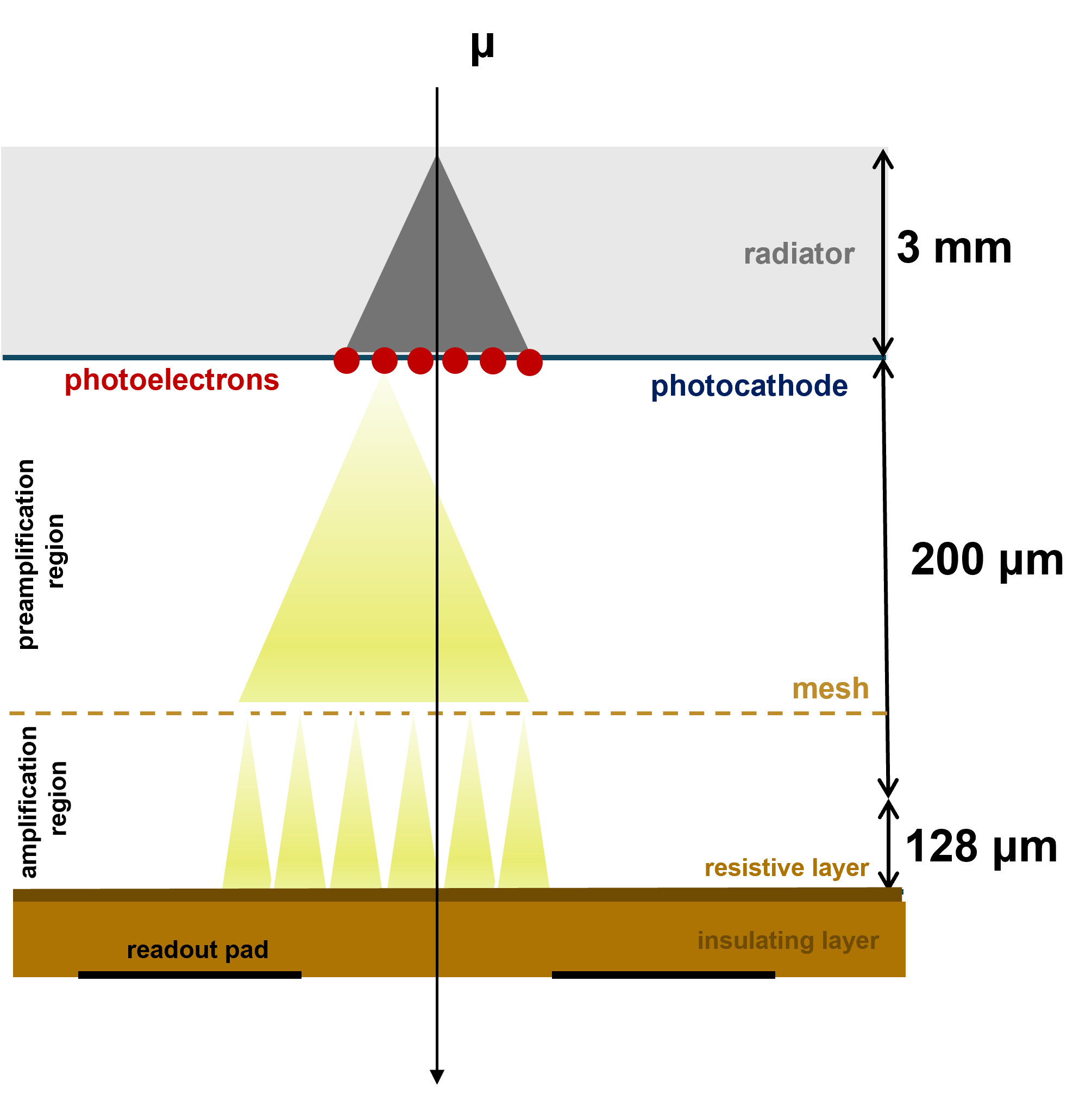}
\caption{\added{Schematic description of the resistive PICOSEC Micromegas detector layout.}}\label{fig:picosec_schema}
\end{figure}

Since its first proof-of-concept, PICOSEC-MM has achieved remarkable progress in timing performance. (Fig. \ref{fig:picosec_schema}) The initial single-channel prototype, featuring a 1\,cm-diameter active zone and operated in a $\mathrm{Ne}:\mathrm{CF_4}:\mathrm{C_2H_6}$ (80:10:10) gas mixture, employed a 3\,mm $\mathrm{MgF_2}$ radiator with an 18\,nm CsI semitransparent photocathode on a 5.5\,nm Cr substrate. This configuration achieved a record time resolution of 24\,ps with 150\,GeV/$c$ muons and a 75\,ps single-photoelectron resolution~\cite{test}. Subsequent optimization---such as reducing the drift gap from 200\,\textmu m to 120\,\textmu m---further improved the single-photoelectron resolution to 50\,ps. More recent refinements in signal integrity, noise reduction, and mechanical reassembly have pushed the timing resolution down to 13.3\,ps~\cite{utrobicic2024singlechannelpicosecmicromegas}, confirming the exceptional potential of this technology.

An excellent candidate to demonstrate the application performance of PICOSEC-MM is ENUBET (Enhanced NeUtrino BEams from kaon Tagging), which aims to produce a narrow-band neutrino beam with precise knowledge of the neutrino flux, flavor, and energy~\cite{longhin2022enhancedneutrinobeamskaon}. The ENUBET beamline is designed to monitor the charged leptons produced in kaon and pion decays within an instrumented decay tunnel. Achieving event-by-event correlation between neutrinos observed in the far detector and the associated charged leptons at the source requires sub-nanosecond timing precision. Here, PICOSEC-MM detectors can play a crucial role.

\begin{figure}[hbt!]
 \centering 
 \includegraphics[width=0.8\textwidth]{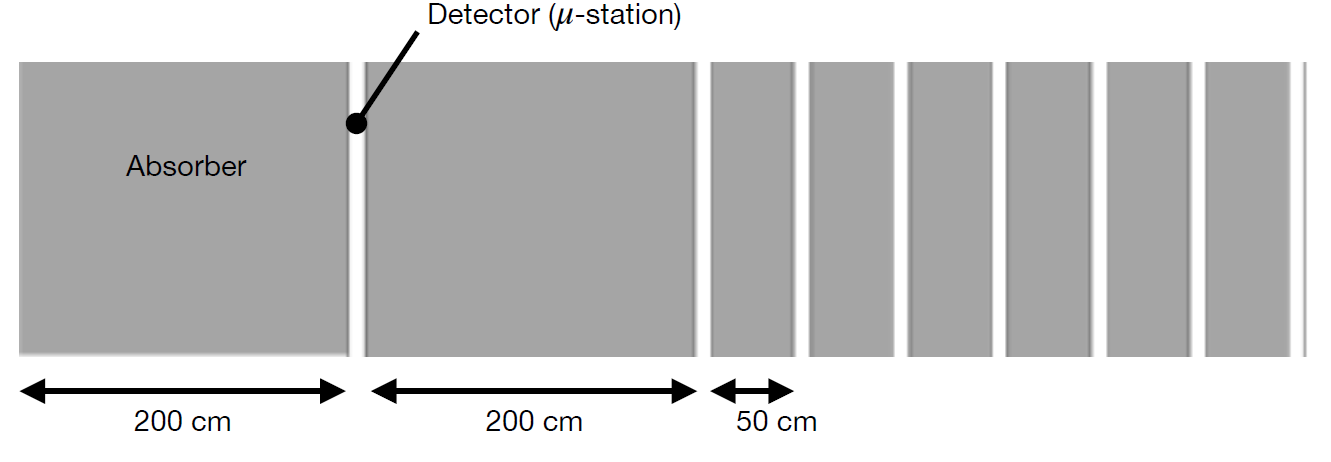}
\caption{Schematic of the muon stations and absorbers configuration to be installed at the end
of the tagger calorimeter. The grey slabs represent the absorbers (made out of iron or rock) while
the white slices, 8 in total, are the muon detector planes \cite{Longhin:2714046}.}\label{fig:ENUBET_absorbers}
\end{figure}

In ENUBET, PICOSEC-MM could be deployed in multiple parts of the facility. \added{In the hadron dump region, the PICOSEC-MM planes integrated downstream of the absorbers could operate as muon monitoring stations, providing precise time tagging of muons originating from pion decays and enabling accurate monitoring of the $\mathrm{\nu_{\mu}}$ flux ~\cite{Longhin:2714046}. A realistic implementation, as can be seen in Fig.~\ref{fig:ENUBET_absorbers}, would consist of two to seven timing planes, each formed by tiles PICOSEC-MM modules, (typical unit size of 100 $\mathrm{cm^{2}}$), covering an active area of the order of 1 $\mathrm{m^{2}}$ to match the transverse muon profile emerging from the dump.} Additionally, PICOSEC-MM modules could serve as timing layers in the electromagnetic calorimeter (EMC) or as a T0 layer in the tagger region, providing prompt time references for electron and positron identification. These applications share standard requirements: large active areas, fine segmentation, radiation tolerance, and long-term operational stability under high particle flux.

To meet these challenges, recent developments have focused on integrating resistive technologies into the PICOSEC-MM architecture \cite{kallitsopoulou:tel-05267379}. Resistive anodes are known to mitigate discharge effects and enhance detector stability in high-rate environments, without degrading timing performance. Previous studies on small-scale, multi-pad prototypes demonstrated that incorporating resistive layers preserves sub-40\,ps resolution while improving robustness and giving insights for the possibility of spatial resolution performance \cite{kallitsopoulou:tel-05267379}. Building upon these results, the next step is to extend the concept to larger active areas suitable for deployment in ENUBET’s muon stations, as presented in the different scenarios previously.

This work presents the design, construction, and performance characterization of a large-area, resistive PICOSEC Micromegas detector developed specifically for ENUBET applications. The new prototype aims to validate scalability while maintaining precise timing, respecting uniformity, and adhering to the low material budget goals set by ENUBET. The following sections describe the detector architecture, the readout implementation, and the results of timing performance measurements obtained in test beams.

\section{The 96-pad Resistive Prototype Development}

\subsection{Design of the Micromegas Board and Spacer}

The development of the new multi-pad PICOSEC Micromegas board was guided by the specific operational and geometrical requirements of the ENUBET experiment. For the first time, a prototype was conceived from the outset to address application-driven constraints rather than adapting an existing design. This represents a significant engineering challenge, as it requires maintaining the mechanical flatness and drift gap uniformity demonstrated in previous large-area prototypes~\cite{Utrobicic_2023}, while ensuring compatibility with dedicated front-end electronics capable of sustaining ENUBET’s data acquisition rates without compromising timing performance.
\begin{figure}[hbt!]
\centering 
\includegraphics[width=0.8\textwidth]{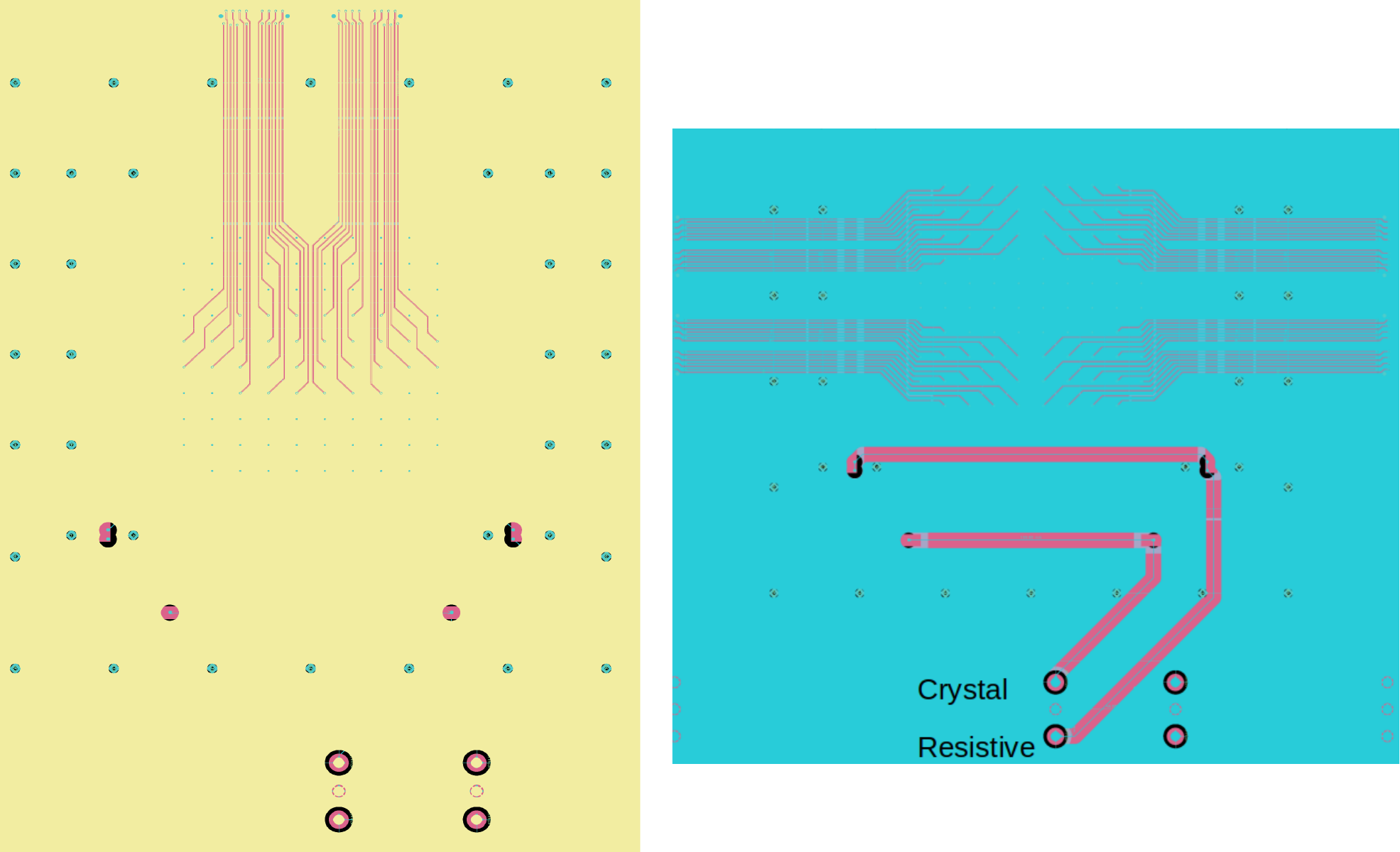}
\caption{Signal routing configuration for the 96-pad board. Two internal copper layers were used to balance signal length and minimize crosstalk.}
\label{fig:96-pad-signal-lines}
\end{figure}

Two versions of the Micromegas board were fabricated using FR4 printed circuit boards of 1.6\,mm and 0.8\,mm thickness, each comprising six copper layers. This dual approach was adopted to assess the influence of substrate thickness on mechanical planarity. The signal lines were routed symmetrically toward the board edges, distributed over two inner copper layers separated by a grounded plane to reduce crosstalk and equalize signal propagation paths, as shown in Fig.~\ref{fig:96-pad-signal-lines}. The 96 readout channels were grouped into four sets of 16 pads, routed to three sides of the detector. Each side hosts 32 channels interfaced through pairs of SAMTEC\footnote{QRM8-026-05.0-L-D-A-GP} connectors for compact, low-impedance readout.

\begin{figure}[hbt!]
\centering 
\includegraphics[width=0.9\textwidth]{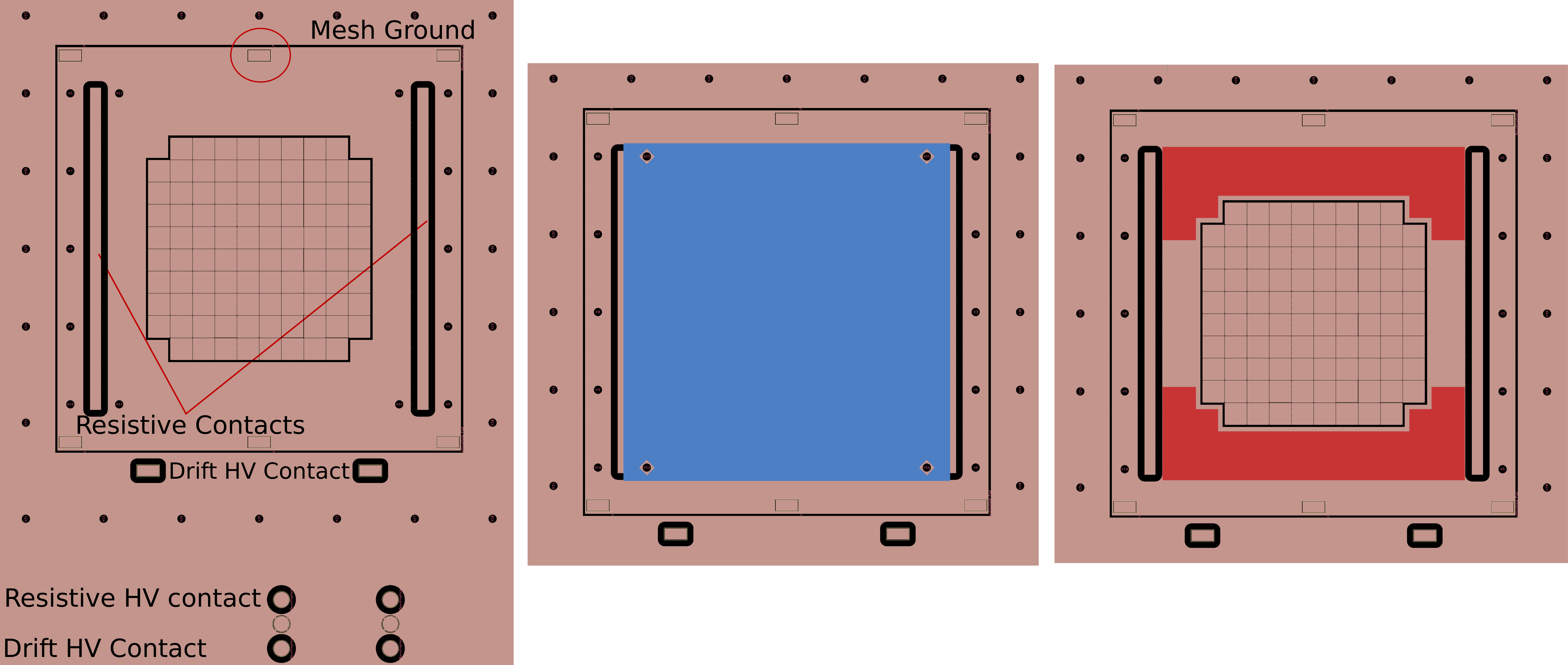}
\caption{\added{Left: Readout plane of the 96-pad detector design. The readout PCB comprises 96 channels with a pad size of $1,\mathrm{cm}^2$. Electrical contact to the resistive foil is achieved along the sides using resistive paste. The mesh is cut beyond the active area and features six grounding points to ensure symmetric charge evacuation across the active region. Middle: Resistive diamond-like carbon (DLC) layer deposited on top of the readout pads. Right: Layout of the drift spacers on the PCB plane. The corner pads are grounded and serve to facilitate the high-voltage connection to the photocathode positioned above the spacers.}}
\label{fig:96-pad-drawing}
\end{figure}

The topmost copper layer of the board forms the readout matrix, consisting of 96 square pads, each 1\,cm $\times$ 1\,cm. The four corner pads are grounded, and the entire pad array is surrounded by a grounded perimeter frame. A stainless steel micromesh is embedded between two laminated polyimide coverlays. The lower (127\,\textmu m thick) coverlay includes openings for the active region, gas circulation, and mesh-to-ground connection, while the upper (50\,\textmu m thick) coverlay mirrors these openings and defines the mesh boundary at the detector’s active edge (see Fig.~\ref{fig:96-pad-drawing}). This configuration ensures a mechanically stable mesh surface and uniform drift gap across the full active area.

The drift spacer plays both a mechanical and electrical role within the assembly. It defines the drift gap thickness while providing the high-voltage (HV) connection to the photocathode. To simplify construction and ensure robust assembly, the spacer was realized as a 50\,\textmu m-thick, three-layer copper–polyimide laminate. Copper traces on its upper face are patterned to align with the grounded corner pads of the readout plane (Fig.~\ref{fig:96-pad-drawing}), which provide the only contact points for grounding. This design minimizes the likelihood of discharges, especially in high-field regions near mesh edges or conductive epoxy junctions.

A $\mathrm{MgF_2}$ crystal coated with a photocathode and equipped with a conductive ring is mounted above the spacer. The ring provides the electrical bias to the photocathode, connected to the HV line via dedicated reference contacts on the PCB. This compact integration ensures a stable electrical interface and reproducible mechanical alignment during assembly.

\subsection{Design of the Detector Chamber}

The detector chamber consists of a rectangular aluminum housing that provides both mechanical support and environmental protection for the active volume. A non-conductive internal insert, fabricated by high-precision resin-based 3D printing, supports the $\mathrm{MgF_2}$ crystal and the spacer assembly. This material was selected as an alternative to PEEK due to its comparable mechanical rigidity, electrical insulation properties, and low outgassing rate. Precision-machined grooves on both sides of the aluminum body accommodate O-ring seals, ensuring gas-tight operation.

Assembly proceeds from the front side of the chamber. The Micromegas board is first positioned on the aluminum housing, followed by the placement and soldering of the spacer. A peripheral support frame is then installed to ensure the precise centering of the $\mathrm{MgF_2}$ window. A dedicated top component is fastened above the crystal using low-torque screws, applying uniform pressure without inducing mechanical stress or deformation.

\begin{figure}[hbt!]
\centering 
\includegraphics[width=0.8\textwidth]{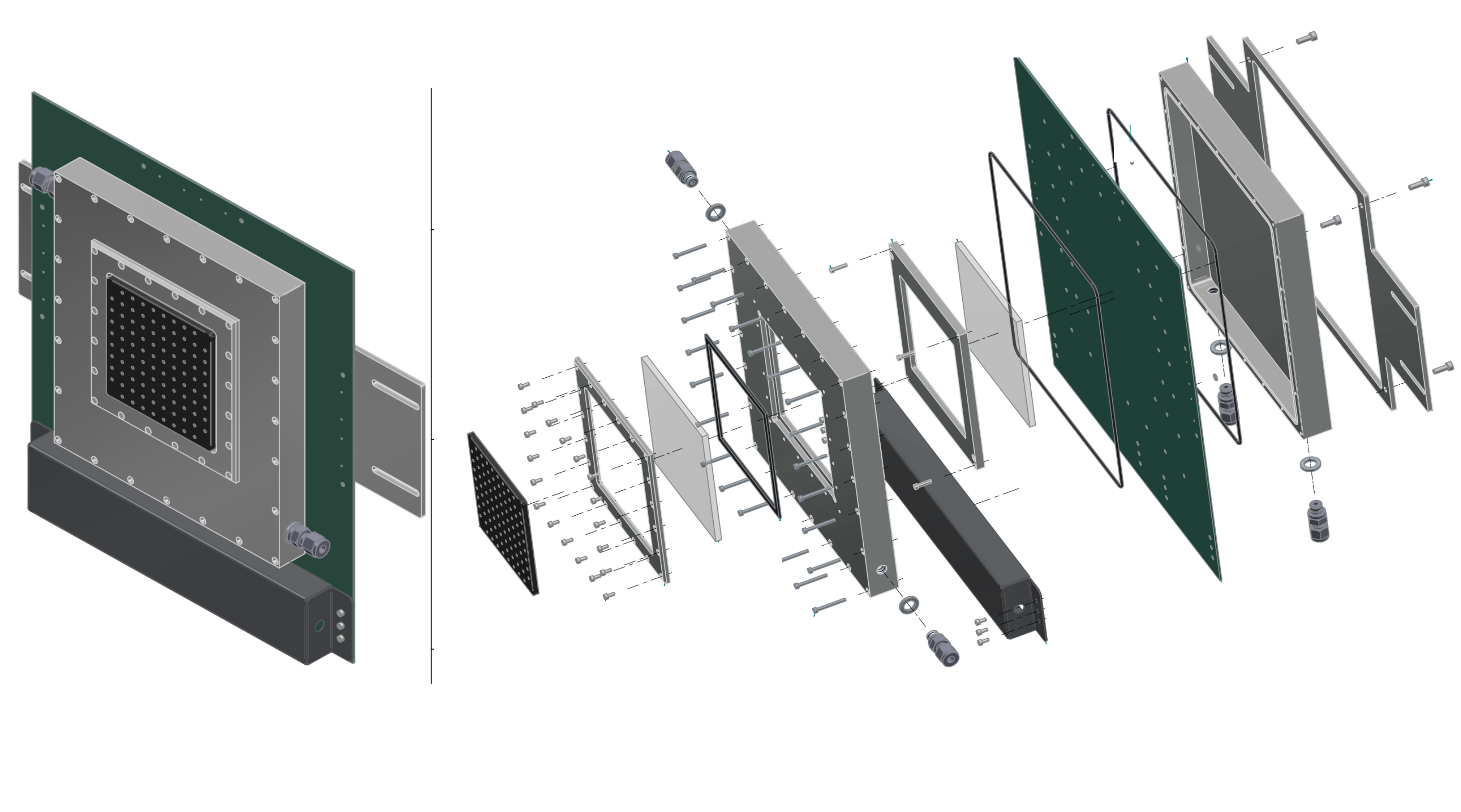}
\caption{\added{Exploded view of the 10$\times$10 pad detector assembly. From left to right: 3D-printed entrance window for single-photoelectron calibration, 3\,mm-thick quartz entrance window with its support and flange ensuring gas tightness and vacuum compatibility, front aluminum window of the entry chamber, spacer defining the drift gap and HV on the radiator, radiator, O-ring for gas tightness, bulk-Micromegas PCB, rear O-ring and aluminum backplate, and finally the metallic support used to mount the chamber on the test-beam telescope. }}
\label{fig:96-pad-mechanics}
\end{figure}

The front of the chamber is closed with a machined aluminum flange incorporating a 3\,mm-thick square quartz window. This window seals the gas enclosure while enabling the transmission of UV light into the detector volume, allowing for controlled calibration measurements. An exploded view of the complete 96-pad detector chamber assembly is shown in Fig.~\ref{fig:96-pad-mechanics}.

\subsection{Planarity Measurements}

Achieving and maintaining excellent planarity is a key mechanical requirement for this detector generation, as it directly affects the uniformity of the drift and amplification gaps, and consequently the time resolution.

In previous studies \cite{AUNE2021165076}, it has been observed that residual deformation across the active area—approximately 200\,\textmu m between the detector center and its periphery—has a measurable impact on timing performance. This sag corresponds to a local variation in the drift gap of roughly 3\%, which translates into a comparable fractional change in the electron transit time to the amplification region. The effect becomes evident in the timing uniformity across the pad matrix.

In this analysis, two distinct configurations for Minimum Ionizing Particles (MIPs) traversing the detector were examined. In the first case, where the pad signal is induced by the majority of the photoelectrons generated by the Cherenkov emission, the time resolution of the central pad—quantified by the RMS spread of the Signal Arrival Time (SAT) distribution—reaches $(26.5 \pm 0.5)$\,ps. In the second case, where the signal originates from only a fraction of the available photoelectrons, the central pad resolution is $(34.0 \pm 0.6)$\,ps. The peripheral pads exhibit systematically worse performance, averaging $\sim33.5$\,ps and $\sim49$\,ps in the same two conditions, respectively. 

These results confirm that local deviations in the drift gap induced by mechanical deformation directly degrade the uniformity of the time response. Achieving sub-10\,\textmu m planarity over large areas is therefore essential for preserving homogeneous timing performance across the full detector surface, especially in large-scale systems such as those envisioned for ENUBET. Given the typical electron drift velocity in the preamplification region ($v_d \approx 10^6$\,m/s), a 10\,\textmu m variation in the gap corresponds to a timing spread of about 10\,ps, establishing the mechanical flatness tolerance required to maintain a uniform sub-30\,ps time resolution.

The Micromegas boards were manufactured using standard FR4 substrates and the bulk Micromegas process, during which each PCB was pressed against a precision marble table to preserve surface flatness. To ensure the required mechanical precision, a comprehensive quality assurance and control (QA/QC) procedure was implemented throughout production. Planarity measurements were systematically performed at multiple stages—before and after each fabrication step—to identify and quantify sources of deformation. Measurements were conducted using a Mitutoyo coordinate measuring microscope~\footnote{\added{\url{https://measure.mitutoyo.com/qv-active-promo/}(model: QV-X404P7L-D)}}.

Initial baseline measurements were taken on unprocessed PCBs to evaluate their intrinsic flatness. Each board was scanned over a dense grid of measurement points, and the root-mean-square (RMS) surface deviation was calculated. The results, indicate that the thick (1.6\,mm) boards exhibited surface variations of approximately 25\,\textmu m, while the thinner (0.8\,mm) boards achieved slightly better planarity, around 15\,\textmu m.

\begin{figure}[hbt!]
\centering 
\includegraphics[width=1.0\textwidth]{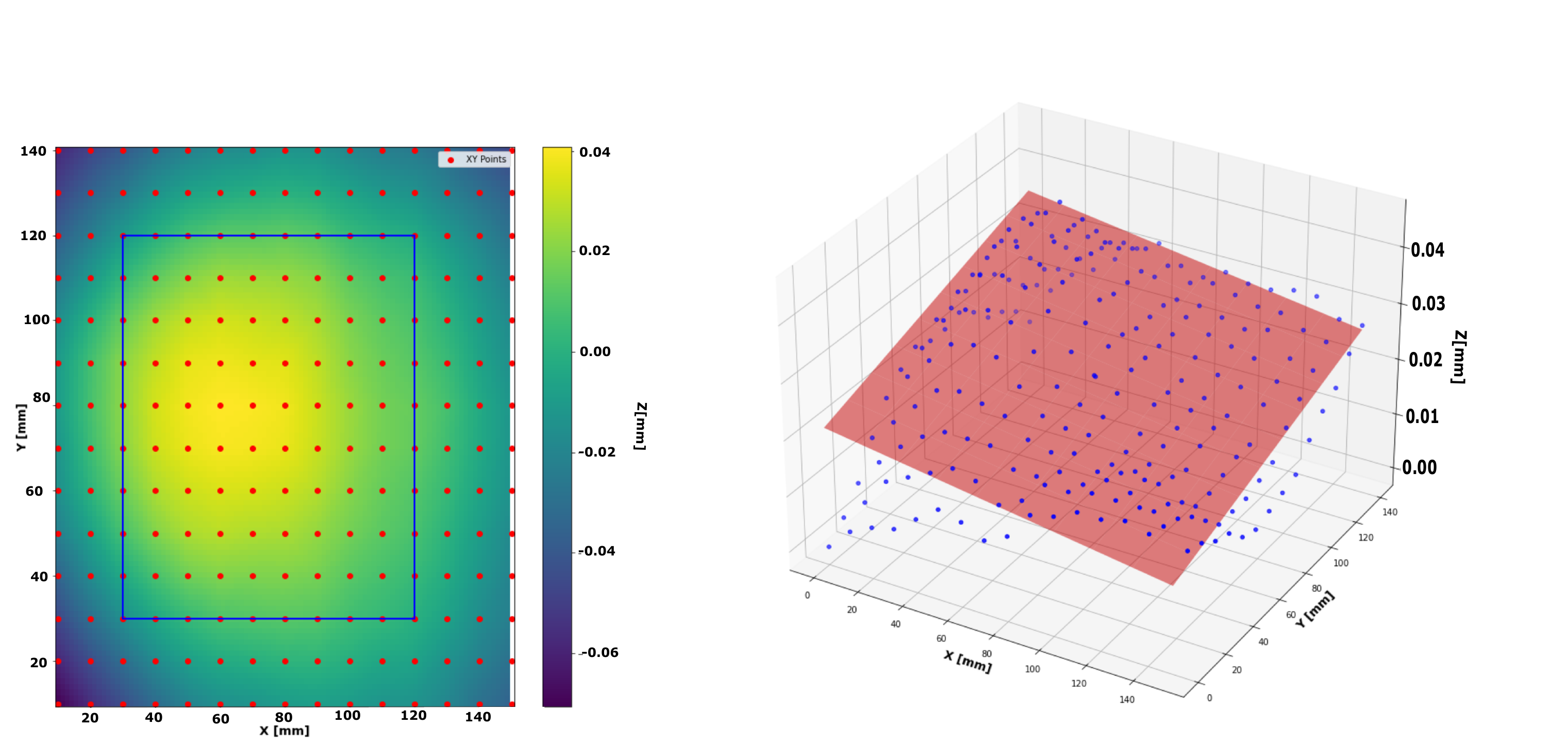}
\caption{Example of a planarity measurement for a thick PCB of 1.6\,mm thickness. Left: Measurement points on the raw PCB plane. With the red points, the points of the grid, used for the scan, and with the blue square, the active zone region. Right: Fit plane on the measured points, resulting in a standard deviation of 20\,\textmu m.}
\label{fig:planarity-thick-board-dlc}
\end{figure}

The most significant deformations were observed after pressing the resistive Diamond-Like Carbon (DLC) film to the Kapton foils, which were laminated onto the PCB prior to the bulk process. The mechanical stress introduced by this step—particularly for films with surface resistivity of 10\,M$\Omega$/$\square$ —resulted in planarity deviations of up to 122\,\textmu m (see Fig.~\ref{fig:planarity-thick-board-dlc}). Such deviations, if uncompensated, could lead to non-uniform amplification fields and degraded timing uniformity across the active area.

To mitigate this effect, a lightweight structural stiffener made of Rohacell foam was bonded to the back side of the PCB. Rohacell, a rigid polymethacrylimide-based foam widely used in aerospace and detector construction, offers an excellent stiffness-to-weight ratio and low outgassing~\cite{BOSZE1997224}. The adhesive layer used during bonding compensates for minor surface irregularities, while the Micromegas pillar array and insulating frame ensure that the drift and amplification gaps remain referenced to the precision marble surface. This configuration transfers the flatness of the marble directly to the active detector plane.

\begin{figure}[hbt!]
\centering 
\includegraphics[width=0.8\textwidth]{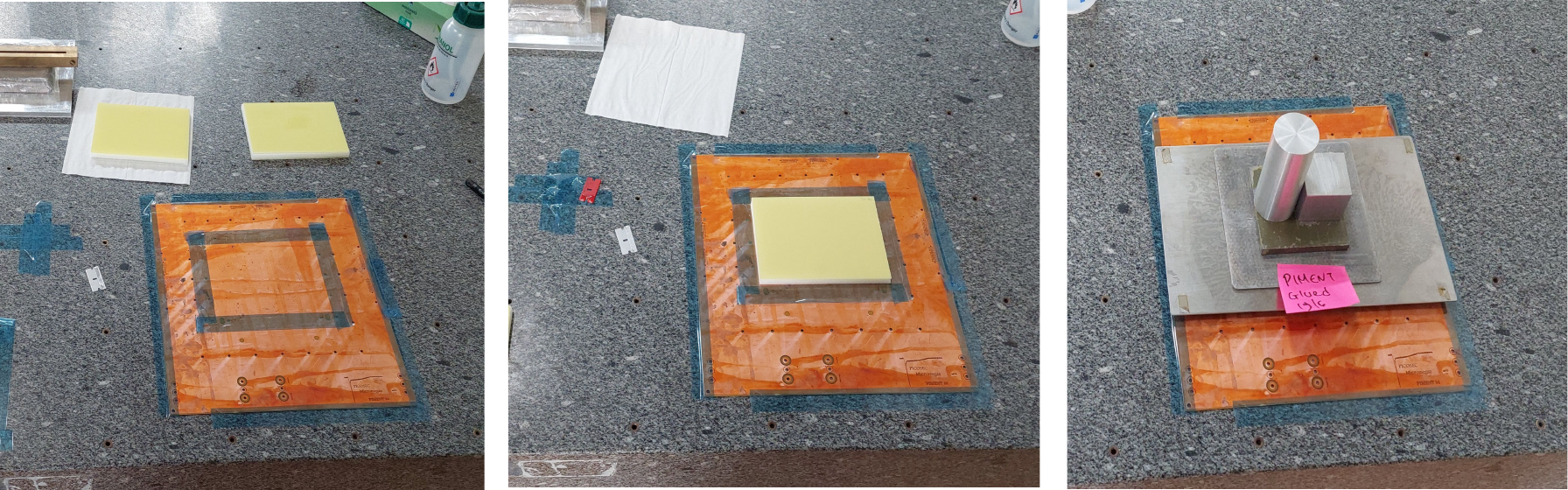}
\caption{Step-by-step gluing of the Rohacell stiffener onto the back side of the bulk PCB.}
\label{fig:96-pad-stiffener}
\end{figure}

The Rohacell stiffener was cut to match the dimensions of the DLC-covered region, thereby counteracting the induced mechanical stress locally. The assembly procedure is illustrated in Fig.~\ref{fig:96-pad-stiffener}, showing the progressive gluing of the support structure.

\begin{figure}[hbt!]
\centering 
\includegraphics[width=1.0\textwidth]{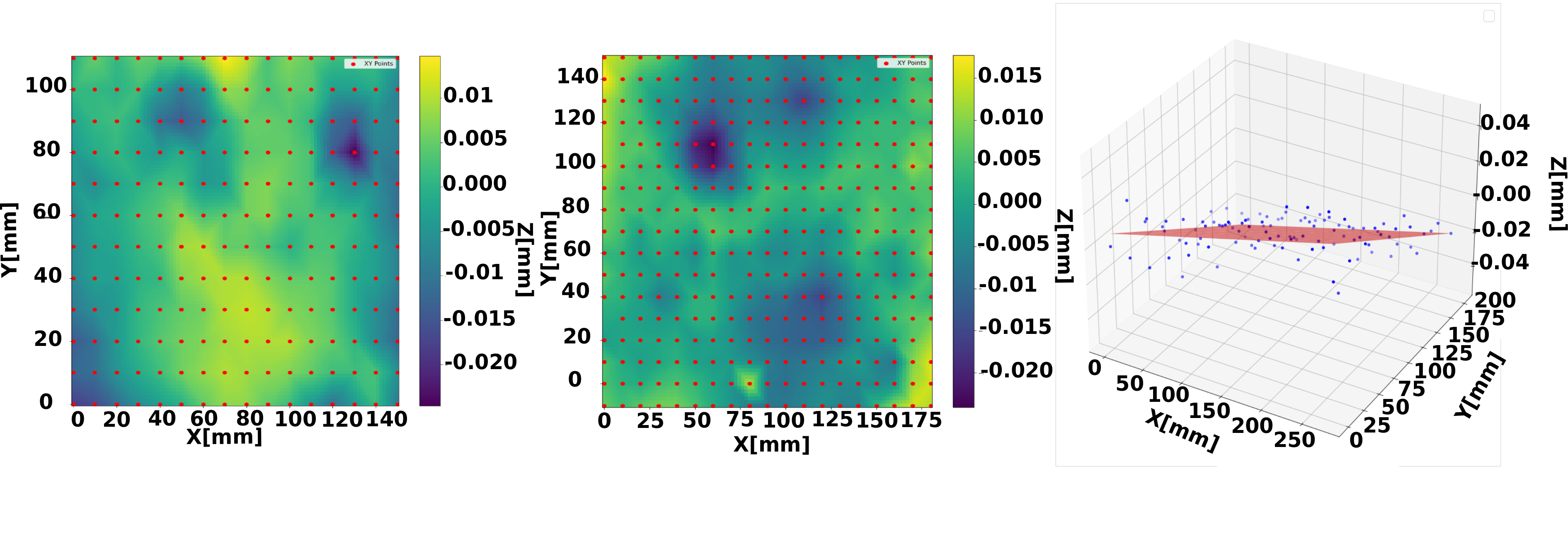}
\caption{\added{Evolution of board planarity through the production process. Left: Planarity of the raw PCB with the resistive plane integrated, approximately around 122\,\textmu m Middle: After bulk Micromegas lamination, RMS deviation is reduced to 68\,\textmu m. Right: 3D plane of the measured points after stiffener gluing, deviation decreases to 10\,\textmu m.}}
\label{fig:96-pad-stiffener-bulking-process}
\end{figure}

Subsequent planarity measurements demonstrated the effectiveness of this approach. The bulk Micromegas lamination process alone reduced the RMS deviation from 122\,\textmu m to 68\,\textmu m, likely due to the tension applied during mesh embedding. After the stiffener was added, the deviation decreased further to approximately 10\,\textmu m, as shown in Fig.~\ref{fig:96-pad-stiffener-bulking-process}. This result confirms that mechanical reinforcement plays a crucial role in maintaining the geometric precision required for uniform field configuration and optimal timing performance.

\section{Experimental Setup and Standard Waveform Analysis}

The measurements presented in this work were carried out during the 2024 RD51/DRD1 collaboration beam test campaigns at CERN. The experimental runs took place at the H4 beamline of the SPS using 150\,GeV muons.

\begin{figure}[hbt!]
 \centering 
 \includegraphics[width=0.8\textwidth]{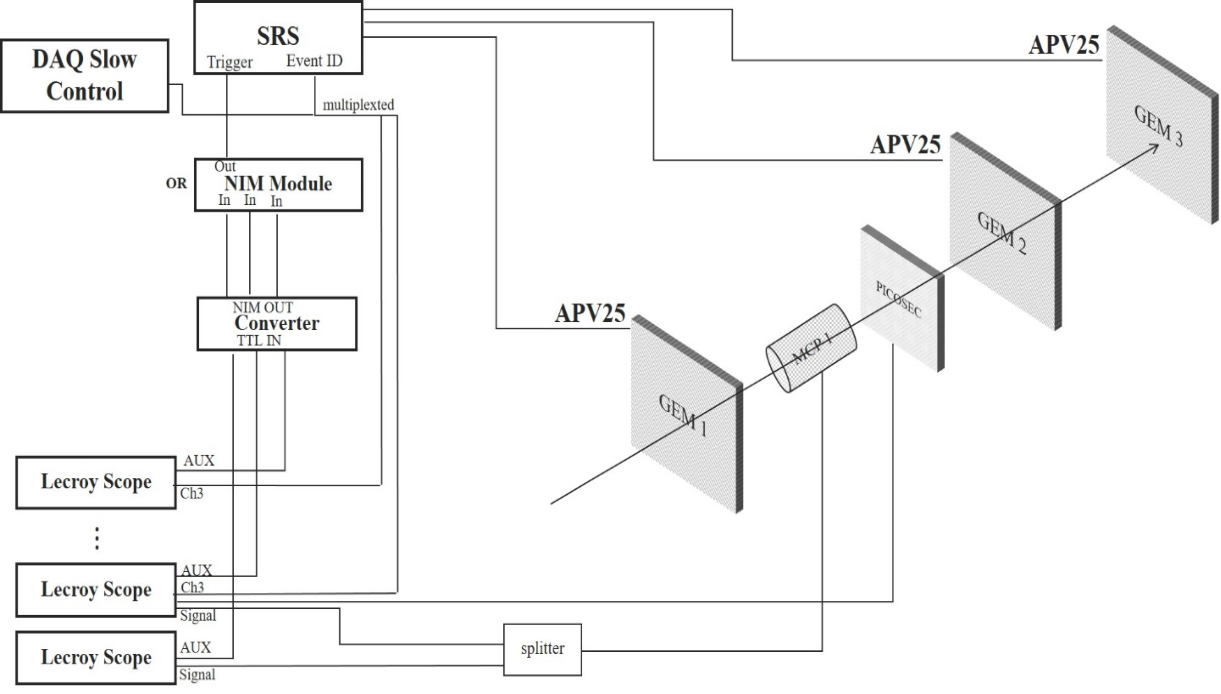}
 \caption{Block diagram and sketch of the electronic modules used to provide NIM signals to trigger the DAQ.}
 \label{fig:triple_gem}
\end{figure}

As illustrated in Fig.\,\ref{fig:triple_gem}, the setup consisted of a beam telescope incorporating the PICOSEC-Micromegas prototype under test, a tracking system made of three triple-GEM detectors, and a Hamamatsu MCP-PMT\footnote{Hamamatsu Microchannel Plate Photomultiplier Tube (MCP-PMT) R3809U-50, \url{https://www.hamamatsu.com/jp/en/product/optical-sensors/pmt/pmt_tube-alone/mcp-pmt/R3809U-50.html}} serving as a precise timing reference. The triple-GEM detectors\footnote{Operated with a 70\% argon and 30\% $\mathrm{CO_2}$ gas mixture at NTP.} were read out using the SRS system~\cite{martoiu2013development}, providing two-dimensional hit reconstruction. Particle trajectories were reconstructed offline using cluster-based hit finding and linear fits, assuming straight paths due to the high beam momentum and absence of magnetic fields~\cite{bortfeldt2014development}.

Trigger signals were generated from the MCP-PMT, converted to NIM logic pulses, and combined through an OR logic condition to form a global trigger. This trigger synchronized the tracker acquisition with the digitized waveforms from the PICOSEC-Micromegas prototype.\added{ The APV25 data \cite{martoiu2013development} were} recorded using the DATE software\footnote{\added{CERN. ALICE DAQ and ECS Manual (2010), \url{https://cds.cern.ch/record/689027/files/INT-1998-04.pdf}.}}, while the PICOSEC-Micromegas waveforms were digitized using Teledyne LeCroy Waverunner oscilloscopes\footnote{Teledyne LeCroy. Waverunner 8000 Series Datasheet (2018)} operated at 10\,GS/s sampling rate.

The detector evaluated in this study was a large-area, resistive PICOSEC-Micromegas prototype equipped with a Diamond-Like Carbon (DLC) layer of surface resistivity \added{10\,M$\Omega$/$\square$}. This configuration was selected following extensive R\&D efforts showing that such resistivity provides the optimal compromise between high-rate stability, discharge mitigation, and sub-40\,ps timing precision \cite{kallitsopoulou:tel-05267379}. The digitized waveforms and associated trigger signals were stored in binary format and processed using a dedicated offline analysis chain.

\subsection{Standard Waveform Analysis}

The offline analysis followed a standardized workflow to ensure consistent signal reconstruction and timing extraction \cite{test}. Each recorded waveform was first corrected for baseline fluctuations, estimated from a pre-trigger window and subtracted event-by-event. Residual high-frequency noise was suppressed using a three-point smoothing algorithm, preserving the integrity of the pulse shape and ensuring reliable charge integration as described in detail in ~\cite{kallitsopoulou:tel-05267379}.

The Signal Arrival Time (SAT) was determined using a Constant Fraction Discrimination (CFD) method applied to a logistic fit of the leading edge of each pulse. The timing was extracted at 20\% of the fitted amplitude, achieving sub-sample precision and minimizing amplitude-dependent biases. The same method was applied to both the MCP-PMT reference and the detector under test. The relative SAT distribution was obtained by subtracting the PICOSEC-Micromegas timing from the MCP-PMT timing, both recorded on the same oscilloscope ~\cite{test}.

Two observables were considered to characterize the pulse shape: the electron-peak charge and the total integrated charge. The electron-peak charge corresponds to the integral of the fast component of the signal, fitted with a double-sigmoid function,
\begin{equation}\label{eq:double_logistic}
    f(x;p_0,p_1,p_2,p_3,p_4,p_5) = p_3 + \frac{p_0}{1+e^{-(x-p_1)p_2}} \times \frac{1}{1+e^{-(x-p_4)p_5}},
\end{equation}
where $p_0$ and $p_3$ are the amplitude limits, and $(p_1,p_2)$ and $(p_4,p_5)$ describe the rise and fall edges, respectively. In practice, the total integrated charge—evaluated up to the local minimum of the cumulative waveform within a 200\,ns window—proved to be a more robust observable as used in similar analyses as in ~\cite{kallitsopoulou:tel-05267379}.

For the data taking concerning the prototype characterization, the MCP-PMT was aligned with the pad under study of the reference PICOSEC-Micromegas detector. Only the central region of the MCP, characterized by sub-6\,ps intrinsic timing resolution and uniform response~\cite{Bortfeldt_2020}, was used in the analysis.

In addition to the waveform reconstruction, a precise spatial alignment of each readout pad was a prerequisite for the accurate evaluation of the detector timing performance. The procedure follows the method developed in our previous work~\cite{kallitsopoulou:tel-05267379}, where the geometrical centers of the pads are determined directly from beam data. For each pad, a two-dimensional charge map was built by correlating the reconstructed track impact points, provided by the GEM telescope, with the total charge measured on that pad. After applying quality filters on track multiplicity and average charge, the charge-weighted projections along the $x$ and $y$ axes were fitted with symmetric second-order polynomials to extract the centroid coordinates $(x_c, y_c)$ of each pad. 

For the 96-pad prototype, this procedure was systematically repeated for all measured channels, yielding a set of calibrated pad centers in the global beam frame. The extracted coordinates were subsequently used to build a transformation matrix that converts the telescope reference system into the detector local frame. This alignment step is a core component of the offline analysis chain, as it directly defines the reference geometry for associating each reconstructed track to the corresponding pad waveform, thus enabling unbiased timing resolution estimation across the detector surface tested.

\subsection{Detector Assembly}

The prototype featured a $10\times10$\,cm$^2$ active area segmented into 96 readout pads, each connected via multi-pin PCB connectors to external electronics. A 2.5\,nm Diamond-Like Carbon (DLC) photocathode with a surface resistivity of 10\,M$\Omega$/$\square$ was employed, following previous R\&D that identified this configuration as optimal for combining high timing performance with robust operation in high-rate or discharge-prone environments~\cite{lisowska2024photocathodecharacterisationrobustpicosec}. The choice of DLC also mitigates the degradation issues previously observed with CsI photocathodes under humid conditions, which has been a core R\&D effort over the past years from the collaboration \cite{lisowska2024photocathodecharacterisationrobustpicosec}.

\begin{figure}[hbt!]
\centering 
\includegraphics[width=1.0\textwidth]{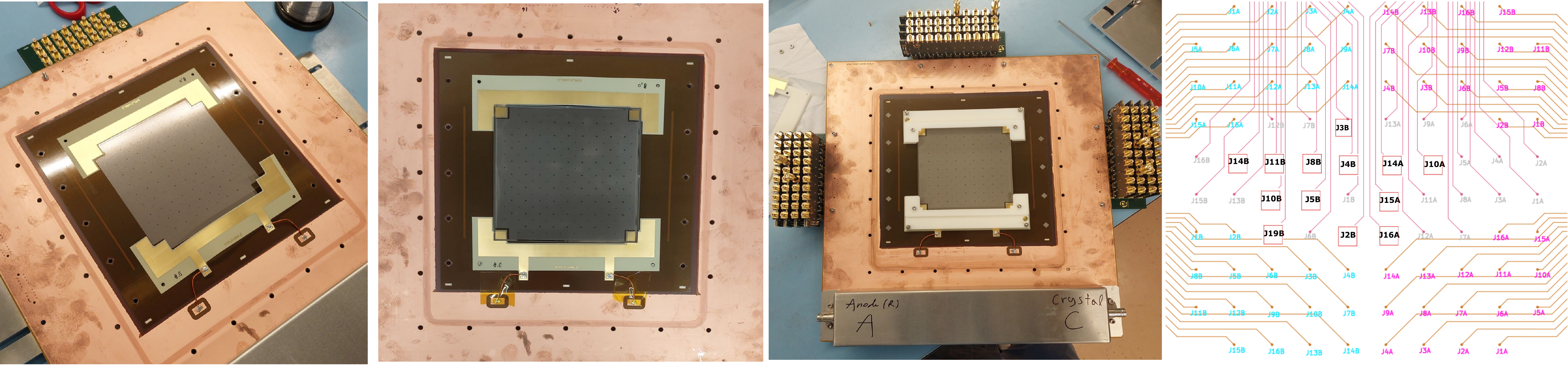}
\caption{Mounting of the 96-pad PICOSEC-Micromegas prototype. The tested pads are indicated in red squares; their relevant positions are centered in the detector.}
\label{fig:96-pad-detector-mounting}
\end{figure}

At the time of testing, the dedicated 16-channel amplifier cards were not yet available. Instead, the detector signals were read out through custom eight-channel preamplifier boards providing 38\,dB gain and 650\,MHz bandwidth, with integrated discharge protection following the design described in~\cite{Hoarau_2021}. A custom PCB adapter converted the SAMTEC connectors to coaxial SMB outputs, allowing waveform acquisition with a 10\,GS/s oscilloscope. Due to the limited number of readout channels, only one-third of the detector was instrumented, and measurements were performed pad-by-pad by repositioning the detector between runs.

\subsection{Beam Test Measurements and Results}

\begin{figure}[hbt!]
\centering 
\includegraphics[width=0.9\textwidth]{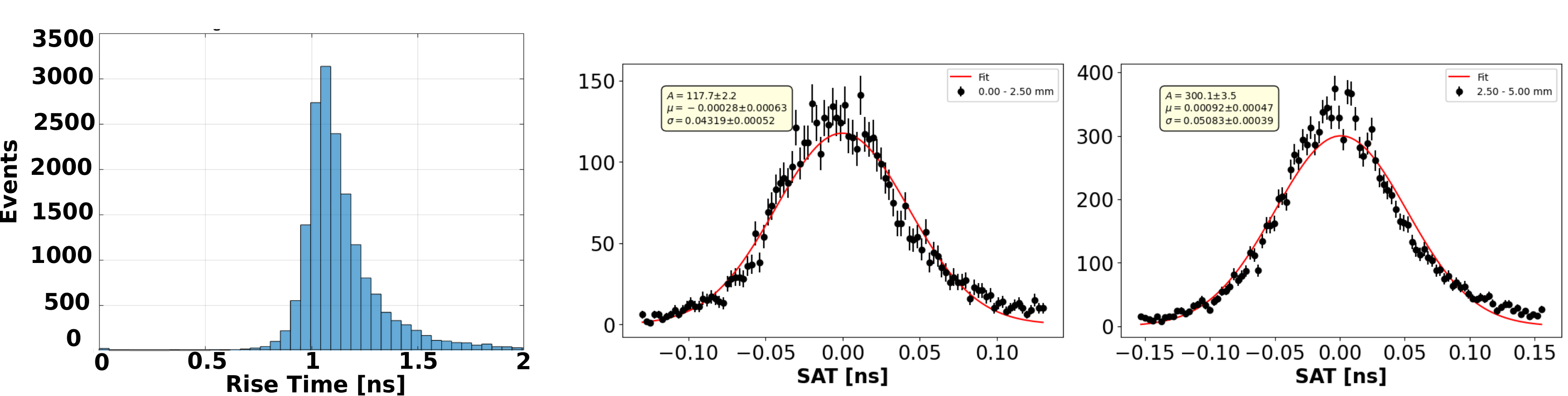}
\caption{\added{(Left) Rise Time distribution. (Middle) Timing resolution of single pads in the central region (0.0 to 2.5\,mm) and (Right) Timing resolution of single pads in a concentric ring of 2.5\,mm to 5.0\,mm radial distance from the pad center. Dashed lines indicate Gaussian fits to the SAT distributions.}}
\label{fig:96-pad-detector-single-pad-timing}
\end{figure}

A voltage scan was conducted to determine the optimal operating point for stable gain and best timing performance. The optimal electric field configuration corresponded to drift and amplification fields of 37.3 and 21.4\,kV/cm, respectively. Under these conditions, the single-pad timing resolution ranged from 43 to 50.1$\pm$0.004\,ps, as shown in Fig.~\ref{fig:96-pad-detector-single-pad-timing}, for the in-pad regions. The mean signal rise time was 1.12\,ns, degraded a bit compared to previously developed prototypes, probably due to the adaptation PCB board on the readout connectors, to match the dedicated amplifier cards. 

\begin{figure}[hbt!]
\centering 
\includegraphics[width=0.8\textwidth]{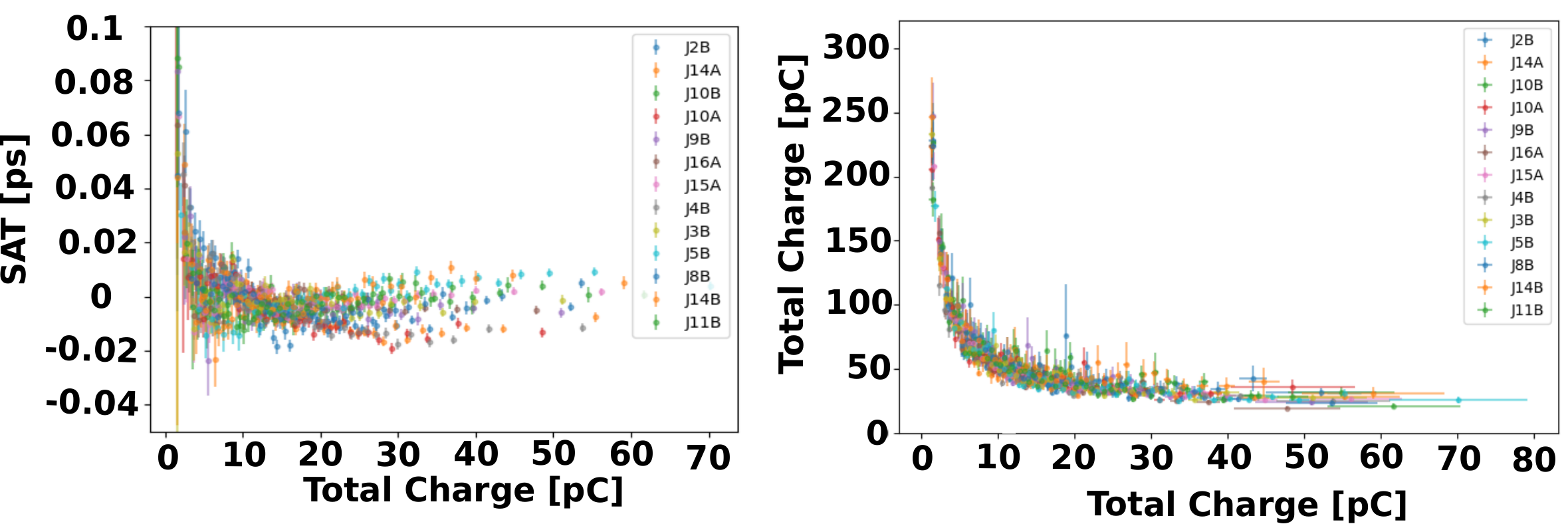}
\caption{\added{(Left) SAT versus total charge parameterization for all tested pads. (Right) Corresponding timing z as a function of total charge.}}
\label{fig:96-pad-detector-SAT}
\end{figure}

Although only a subset of pads could be instrumented simultaneously, all were analyzed individually using the same CFD method as in \cite{kallitsopoulou:tel-05267379}. The SAT and corresponding timing resolution as a function of total charge are shown in Fig.~\ref{fig:96-pad-detector-SAT}. The consistency across pads confirms the intrinsic uniformity of the detector response and the successful effect of the strongback that was used.

\begin{figure}[hbt!]
\centering 
\includegraphics[width=0.9\textwidth]{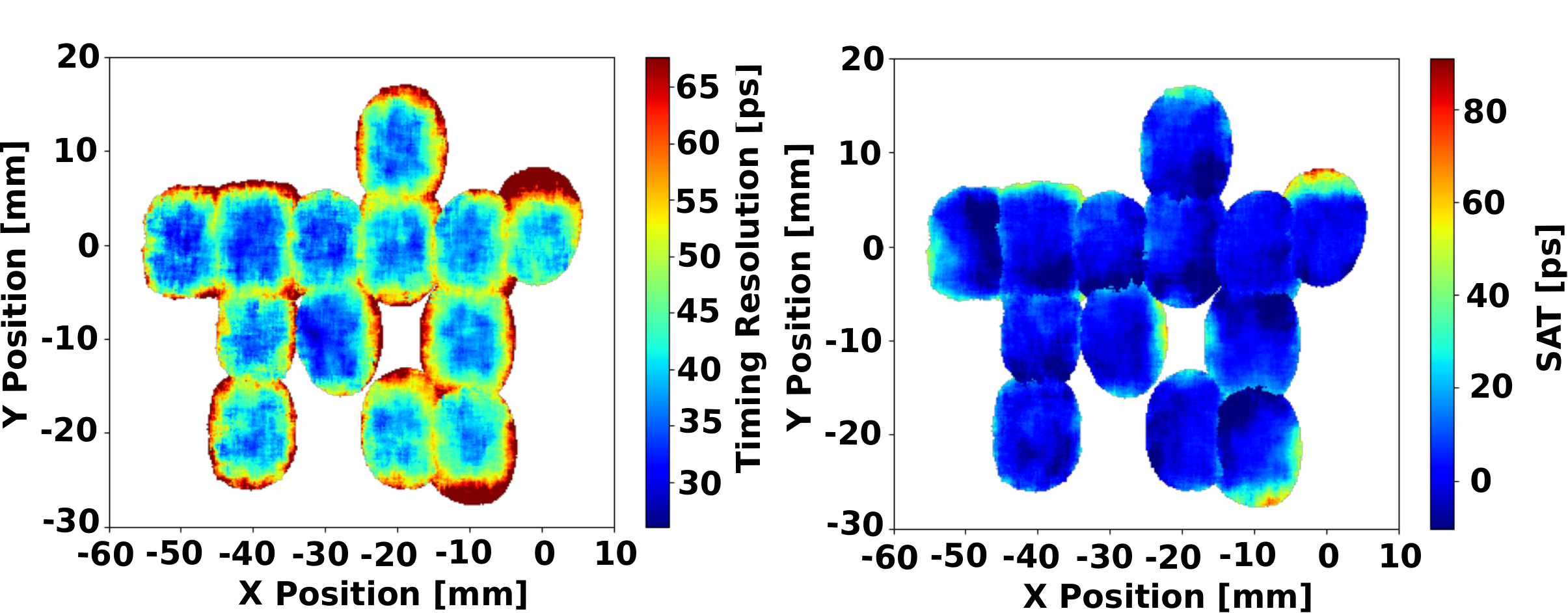}
\caption{\added{Superimposed spatial maps of (left) timing resolution and (right) corrected SAT for all measured pads. Non-instrumented regions appear as empty areas.}}
\label{fig:96-pad-detector-SAT-combined}
\end{figure}

To reconstruct the spatial uniformity, the instrumented area was shifted across successive runs to cover different pad regions. Figure~\ref{fig:96-pad-detector-SAT-combined} shows the superimposed heatmaps of the SAT and timing resolution across the central region of the detector. These maps were obtained by combining the measurements from individual pads, positioned according to their calibrated locations on the detector plane. As only one-third of the 96-pad prototype was instrumented during each run, the detector was repositioned between beam exposures to align the active region with the triple-GEM telescope acceptance. Each data subset was then spatially aligned using the same procedure established in \cite{kallitsopoulou:tel-05267379}, ensuring consistency with the square-pad geometry. The blank areas in the heatmaps correspond to pads that were not equipped with readout electronics during the measurements.

\subsection{Scalability and Future Developments}

The modular architecture of the detector readily supports scaling to active areas up to $30\times30$\,cm$^2$ by tiling $10\times10$\,cm$^2$ $\mathrm{MgF_2}$ crystals. Mechanical design studies indicate that only $\sim4$\% of the total area is lost to inter-crystal dead zones. Future upgrades will implement refined bulk Micromegas structures with extended support pillars and pressure-controlled contact between the mesh and the photocathode to further improve gap uniformity. This configuration is under development and will be evaluated in the next beam campaign planned for 2025.

\section{Conclusions}

The design, construction, and beam characterization of a large-area, resistive PICOSEC-Micromegas (PICOSEC-MM) prototype have been successfully evaluated, demonstrating the scalability and robustness of the technology. The 96-pad detector, featuring a $10\times10$\,cm$^2$ active area and a 2.5\,nm Diamond-Like Carbon (DLC) photocathode with 10\,M$\Omega$/$\square$ surface resistivity, represents a major step toward realizing large-format, picosecond-precision gaseous detectors, designed for applications.

The detector assembly emphasized mechanical planarity and gap uniformity, ensuring sub-10\,µm flatness after lamination with a Rohacell stiffener. This mechanical stability translated into consistent timing performance across the active area, with individual pad resolutions of 43\,ps at the central region of the pads, under 150\,GeV/$c$ muon irradiation. The use of the DLC photocathode provided both high temporal stability and robustness against environmental degradation previously observed with CsI layers.

Despite the temporary limitation of the readout system—restricting simultaneous acquisition to one-third of the pads—the systematic pad-by-pad characterization confirmed the uniformity of the response and the reproducibility of the timing performance over multiple runs. The results validate the effectiveness of the resistive architecture in maintaining high-resolution timing while ensuring protection against discharges.

This work establishes a scalable design methodology for extending PICOSEC-MM detectors to larger active areas. The demonstrated mechanical integrity, electrical stability, and uniform timing response confirm that the adopted architecture meets the requirements of high-rate, high-precision timing applications, such as lepton time-tagging in the ENUBET decay tunnel. The next generation of prototypes, currently under fabrication, will integrate full-channel readout and have improved preamplifier coupling.

In summary, the 96-pad resistive PICOSEC-MM prototype successfully demonstrates that large-area gaseous photodetectors can achieve timing resolutions below 50 ps per pad while preserving robustness and scalability—key milestones toward their deployment in future high-rate timing experiments. The results indicate that the detector concept fully meets the timing and mechanical requirements foreseen for the ENUBET decay tunnel. However, further developments are still ongoing to achieve full system integration, including the completion of the multi-channel readout, long-term stability studies under sustained irradiation, and the optimization of signal coupling to front-end electronics. These next steps will ensure the technology’s readiness for large-scale implementation within the ENUBET framework, in the near future.

\acknowledgments
The authors gratefully acknowledge the support of the French National Research Agency (ANR) through the project \textit{“Development of a PICOSEC-Micromegas detector for ENUBET – PIMENT
”}(ANR-21-CE31-0027). The authors also acknowledge the support of the RD51 collaboration in the framework of RD51 common projects, as well as the French Alternative Energies and Atomic Energy Commission (CEA). 
This work was partially supported by the European Union’s Horizon 2020 research and innovation program through the STRONG-2020 project under grant agreement No.~824093.

\bibliographystyle{JHEP}
\bibliography{main}
\end{document}